\documentclass[10pt]{llncs}

\usepackage{epsfig}

\usepackage{amssymb}
\usepackage{mathrsfs}

\usepackage[titletoc]{appendix}

\usepackage{url}
\urldef{\mailsa}\path|{pankaj.dayama, aditya.karnik}@gm.com|
\urldef{\mailsb}\path| hari@csa.iisc.ernet.in|

\def\QED{\mbox{\rule[0pt]{1ex}{1ex}}}

\usepackage[square, numbers]{natbib}
\makeatletter
\renewcommand\bibsection
{
	\section*{\refname
	\@mkboth{\MakeUppercase{\refname}}{\MakeUppercase{\refname}}}
}
\makeatother

\begin{document}

\title{Optimal Mix of Incentive Strategies for Product Marketing on Social Networks}

\author{Pankaj Dayama\inst{1}, Aditya Karnik\inst{1} \and Y. Narahari\inst{2}}

\institute{Global General Motors R\&D, India Science Lab, GM Technical Center (India), Bangalore 560066, India \\
\mailsa\\
\and Department of Computer Science and Automation, Indian Institute of Science, Bangalore 560012, India\\
\mailsb}

\maketitle 
% use optional labels to link authors explicitly to addresses:
% \author[label1,label2]{<author name>}
% \address[label1]{<address>}
% \address[label2]{<address>}
% \author{Pankaj Dayama\corref{cor1}}
% \ead{pankaj.dayama@gm.com}

% \author{Aditya Karnik\corref{}}
% \ead{aditya.karnik@gm.com}

% \cortext[cor1]{Corresponding author}

% \address{General Motors Global R\&D India Science Lab, International Tech Park, Bangalore, India 560 066}

% \begin{center}
% {\bf Optimal Mix of Marketing Strategies} \\

% \end{center}

% \begin{keyword}
% keywords here, in the form: keyword \sep keyword

% Marketing strategies \sep Social networks \sep Mobile wireless networks

% MSC codes here, in the form: \MSC code \sep code
% or \MSC[2008] code \sep code (2000 is the default)

% \end{keyword}

%\end{frontmatter}

% \section*{Summary of Results}
% We consider the problem of optimally rolling out two incentive strategies, viz., direct incentives and referral rewards, over a finite time horizon $T$. We show that the optimal strategy for the seller has a simple structure where the seller needs to run each of the schemes at most twice for a certain duration. The overall optimal strategy is non-adaptive (or open-loop). This simplifies implementation and practically can help a seller pre-allocate budget for its campaign.
%  Through numerical examples, we also show that when the seller has a high level of past reputation, `exploit-and-influence' strategy is optimal. On the other hand, `influence-and-exploit' strategy is optimal when social influence has a strong bearing on the population.  
% \newpage

\begin{abstract}
%\section*{Summary}
We consider the problem of devising incentive strategies for viral marketing of a product. In particular, we assume that the seller can influence penetration of the product by offering two incentive programs: a) direct incentives to potential  buyers ({\em influence}) and b) referral rewards for customers who influence potential buyers to make the purchase ({\em exploit connections}). The problem is to determine the optimal timing of these programs over a finite time horizon. In contrast to algorithmic perspective popular in the literature, we take a mean-field approach and formulate the problem as a continuous-time deterministic optimal control problem. We show that the optimal strategy for the seller has a simple structure and can take both forms, namely, {\em influence-and-exploit} and {\em exploit-and-influence}. We also show that in some cases it may optimal for the seller to deploy incentive programs mostly for low degree nodes. We support our theoretical results through numerical studies and provide practical insights by analyzing various scenarios.  

%\category{J.4}{Social and Behavioral Sciences}{Economics}

%\category{H.4}{Information Systems Applications}{Miscellaneous}

%A category including the fourth, optional field follows...
%\category{D.2.8}{Software Engineering}{Metrics}[complexity measures, performance measures]

%General terms should be selected from the following 16 terms: Algorithms, Management, Measurement, Documentation, Performance, Design, Economics, Reliability, Experimentation, Security, Human Factors, Standardization, Languages, Theory, Legal Aspects, Verification.

%\terms{Economics, Theory, Performance}

\end{abstract}

\section{Introduction}
\label{sec:intro}
A key research topic in multi-agent systems is to understand the effect of microdynamics/interactions between agents on macroscopic properties. Often the agents are a part of a social structure such as a social network. A common example is that of a social network that consists of potential buyers of a particular new product offering in the market. These buyers interact with each other and influence each others' purchase decisions through word-of-mouth and/or behavior. This so-called {\em social influence} exerted by agents on their neighboring agents in the network have a significant role to play in generating a network effect on the sales of the product. The idea of viral marketing is to essentially exploit the (macroscopic) network effects that result due to the microdynamics between the agents in the network.

Viral marketing is receiving much attention by practicing marketers and academics alike. While not a new idea, it has come to the forefront because of multiple effects - products have become more complex, making buyers to increasingly rely on opinions of their peers; consumers have evolved to distrust advertising; and Web2.0 has revolutionized the way people can connect, communicate and share. With power shifting to consumers, it has become important for sellers to devise effective viral marketing strategies  \citep{godes-etal05}. This work is motivated by this urgent need.

For social influence to work, there must be {\em seeds}, i.e., product advocates to start with. The sellers, therefore, employ two basic strategies. The first is to create advocates, by providing incentives to potential buyers to make an actual purchase. These incentives are typically in the form of discounts, free goodies, etc. The second is to reward product advocates who `put in a good word' and influence potential buyers to make the purchase. Thus, the latter program helps to {\em exploit} the impact of social influence while making a purchasing decision whereas the former program helps to directly {\em influence} the buying behavior by offering discounts.

Since incentives come at a cost, a seller must balance the revenue she generates through these strategies and the expenditure she incurs in doing so. This poses some non-trivial challenges. The first is determining incentives themselves, since response of an individual is contingent on them (too low a referral reward may not elicit recommendation from an individual since personal reputation is usually at stake). Secondly, the two programs are not necessarily causally connected. The reputation of a firm or a brand might create product advocates without incentives, thereby, requiring a seller to launch a referral program directly. This necessitates careful `timing' of these programs.

The objective of this paper is to shed some light on this practically important and theoretically interesting problem. In particular, we seek to determine an optimal timing of these programs over a finite time horizon.

\subsection{Related Work}
In recent years, problems such as these have attracted much attention. Several papers investigate `influence maximization' (see, for example, \citet*{Domingos01,Kempe03,Bharathi07, Chen10}), where the problem is to determine the set of initial adopters who, through an influence process, can maximize the future adoptions of the product. \citet*{Auriol00} discuss a dynamic model of how standards and norms emerge in decentralized economies. \citet*{Hartline08,Arthur09} consider the problem of `revenue maximization' for viral marketing and are close in spirit to the problem we consider in this paper. 

In \citet{Hartline08} a model is proposed in which the purchase decision of a buyer is influenced by individuals who own the product and the price at which the product is offered. An optimal pricing policy is derived using dynamic programming in a symmetric setting (i.e., identical buyers). In a general setting, finding an optimal strategy is shown to be NP-hard and approximation algorithms are considered. The authors suggest {\em influence-and-exploit} strategy where selected buyers are given the product for free, and the seller extracts revenue by making a random sequence of offers and a greedy pricing strategy for the remaining buyers to compensate for the initial loss.   

\citet{Arthur09} also considers a model in which a buyer's decision is influenced by friends who own the product and price at which the product is offered. Sales are assumed to cascade through the social network. The seller offers cashback to recommenders and also sets price for each buyer. The authors show that determining an optimal strategy to maximize expected revenue is NP-hard and propose a non-adaptive {\em influence-and-exploit} policy, which offers product to the interior nodes of the max-leaf spanning tree of the network for free and later exploits their influence by extracting more revenue from the leaf nodes of the tree. They show that the expected revenue generated from the non-adaptive strategy is within a constant factor of the optimal revenue from an adaptive strategy.    

\subsection{Our Contributions}
We consider a seller interested in selling a product to a population of $N$ agents. The product is assumed to be durable and free from network externalities. From the seller's perspective, each agent assumes one of the following types at any point in time: {\em potential buyer} (one who is yet to make a purchase), {\em customer} (one who has purchased the seller's product) and {\em competitor's customer} (one who has purchased a competing product\footnote{All competitors are aggregated into one single virtual competitor.}). A potential buyer makes a purchase decision of her own volition (essentially under {\em external influence}) or under social influence. This decision-making is modeled probabilistically, by specifying for both products (the seller's and the competitor's), probabilities of purchase under external influence and social influence. The seller can influence the former through direct incentives (which affect the price) and the latter through referral rewards. The problem she faces is to roll out these programs so as to maximize the profit, which is equal to the revenue obtained by customer acquisition minus the expenditure on direct incentives and referral rewards, over a given time horizon $T$.

In practice sellers have limited knowledge about the social network underlying the population, typically,  in the form of a {\em class}-level statistical description of it. A class comprises agents who are considered essentially identical on a variety of factors (chosen by the seller), such as demographic, economic level, number of social contacts and so on. In this paper we consider this set-up. However to keep it simple, we assume heterogeneity only in terms of network connections (in particular, probabilities of purchase under either external or social influence are assumed to be the same for all agents); hence classes are based only on the number of social contacts (degrees). The seller thus knows only the degree distribution and degree-degree correlation of the social network. 

%We assume that the seller has limited knowledge about the social network underlying the population. Specifically, she is assumed to have only a {\em class}-level statistical description of it. A class comprises agents who are considered essentially identical on a variety of factors (chosen by the seller), such as demographic, economic level, number of social contacts and so on. To keep the set-up simple, in this paper we assume heterogeneity only in terms of network connections (in particular, probabilities of purchase under either external or social influence are assumed to be the same for all agents); hence classes are based only on the number of social contacts (degrees). The seller thus knows only the degree distribution and degree-degree correlation of the social network.

This class-level statistical description of the agent population allows us to approximate the stochastic evolution of the purchase dynamics by a deterministic process described by ordinary differential equations (ODEs). This is formally established as a mean-field limit, taking the number of agents $N \to \infty$ \cite{Benaim08}. With an ODE limit, we pose the problem as a continuous time optimal control problem and employ the well known Pontryagin's Maximum principle \cite{Kirk70} to characterize an optimal control. An optimal control specifies for each class the times at which direct incentives and referral rewards programs are to be executed. The following are our main results.
\begin{enumerate}
\item We show that an optimal control has a simple structure: the seller needs to run each of the programs at most twice for a certain duration. Moreover, it is non-adaptive (or open-loop). This simplifies the implementation and practically can help a seller pre-allocate the budget for her campaign. 

\item While {\em influence-and-exploit} strategy turns out to be optimal when social influence is strong in the population, {\em exploit-and-influence} strategy can be optimal when the seller has a good reputation.  

\item In some settings, the seller may be better off incentivizing low degree nodes as against the popular approach of targeting the influentials (high degree nodes). This, we believe, provides some support to the findings reported in \cite{Watts07} in reference to the {\em influentials hypothesis}.

\end{enumerate}

The approach we have taken to address the problem is entirely different from the ones in the literature. While a large size of the population presents a challenge to the earlier approaches, it, in fact, aids us in migrating to a simpler deterministic description of the dynamics. The assumption that agents of a class are indistinguishable also fits in naturally with the popular marketing approach of {\em customer segmentation} and allows a seller to customize incentives and referral rewards as per these segments. 

In contrast to earlier papers, we have also modeled competition. This is not only close to reality but interestingly it allows to address some problems in completely different contexts. For example, in limiting the spread of misinformation about an entity or an Internet virus, the objective is to maximize nodes with correct information or security patches by immunizing them (akin to direct incentives) and/or incentivizing them to spread the information they have to their neighbors (akin to referrals). Our results are, thus, applicable to these problems as well (see \cite{Budak11} for discussion of the influence limitation problem).

%%%%%%%%%%%%%%%%%%%%%%%%%%%%%%
\newpage
\section{Problem Formulation}
Consider a population of $N$ agents, indexed by $i=1,2,$ $\ldots,N$. The underlying social network is specified by an undirected graph $\cal{G}=(V,E)$. Each agent is identified with a node in $\cal{V}$ and $(i,j) \in \cal{E}$ means that $i$ and $j$ are social contacts and they influence each other in decision-making. 

$s_i$ denotes the state of agent $i$. $s_i$ can take three values: $0$ (indicates potential buyer), $1$ (indicates customer) and  $-1$  (indicates competitor's customer). Let ${\bf s}:=(s_1,s_2,\ldots,s_N)$.

Each agent makes the purchase decision at a random time point, independent of all others. It suffices to assume that time is discrete (denoted by $n=1,2,\ldots$) and at each time step, an agent is chosen uniformly randomly from the population for a potential state change. Since there are no repeat purchases, $1$ and $-1$ are absorbing states. Therefore, the state change occurs only if the chosen agent is a potential buyer. Suppose agent $i$ is chosen at a time $n$. Then one of the following happens if $i$ is a potential buyer.
\begin{enumerate}
\item $i$ buys the seller's product on her own with probability $\alpha$ (For example, $\alpha=0.08$ means that there is 8\% chance that a potential buyer will buy the seller's product on her own).
\item $i$ buys the competitor's product on her own with probability $\delta$. 
\item $i$ selects one of her social contacts at random. If the selected contact is a customer, $i$ buys the seller's product under social influence with probability $\beta$ (For example, $\beta=0.1$ means that there is  10\% chance that a potential buyer will buy the seller's product if she interacts with someone who has already bought the product).
\item $i$ selects one of her social contacts at random. If the selected contact is a competitor's customer, $i$ buys the competitor's product under social influence with probability $\gamma$.
\end{enumerate}

Clearly, the state process $\{{\bf s}(n),n \ge 1\}$ is a Markov chain. 

Now from the seller's perspective, agents having the same degree are indistinguishable and the network $\cal G$ is known only statistically, i.e., $\cal{G}$ is drawn from an ensemble of random undirected graphs of size $N$, a given degree distribution $P(k)$ ($1 \le k \le K$) and degree-degree correlation function $P(k'|k)$, which denotes the probability that a given link from a node of degree $k$ is to a node of degree $k'$.  Note that a number of well-known graphs such as homogeneous random graphs, exponential random graphs (e.g., $G(n,p)$ and Watts-Strogatz network), scale-free networks can be represented in this framework. We assume that $K$ remains uniformly bounded as $N \rightarrow \infty$.

Denote by $i_k, r_k$ and $\theta_k$ the fraction of degree-k agents who are potential buyers, customers and competitor's customers respectively (note that normalization is with respect to the number of class-k agents; hence $i_k+r_k+\theta_k=1$). Let $x_k:=(i_k, r_k,\theta_k)$ and ${\bf x}:=(x_1,\ldots,x_K)$. From the above assumption, it follows that $\{{\bf x}(n),n \ge 1\}$ is a Markov chain (as seen by the seller). 

The drift of ${\bf x}$ can be computed considering the four cases described above. Table~\ref{tab:tran1} shows the corresponding probabilities and the change in $x_k$ for degree class-k. Consider as an example Case $3$. The probability of a randomly selected agent being a potential customer of degree $k$ is $P(k)i_k$. This agent randomly chooses one of her $k$ social contacts. The probability that this chosen one is an existing customer is 
$\scriptstyle{\frac{1}{k}\sum_{j=1}^k \sum_{k' \in K} {P(k'|k) r_{k'}}}$ = $\sum_{k' \in K} {P(k'|k) r_{k'}}$. 
The selected agent buys the seller's product under the social influence from her contact with probability $\beta$. Thus the probability of Case 3 is $\beta P(k)i_k \sum_{k' \in K} {P(k'|k) r_{k'}}$.  One agent changes her state from $0$ (potential customer) to $1$ (customer). Hence the effect on $x_k$ is $\frac{1}{NP_k}(-1,1,0)$. %(note that $N_k=NP(k)$).

\begin{table}[b]
\begin{center}
\begin{tabular}{l|l|l}
\hline
Case & Probability & Effect on $x_k$ \\ \hline
1 & $\alpha P(k)i_k$ & $\frac{1}{NP_k}(-1,1,0)$\\
2 & $\delta P(k)i_k$ & $\frac{1}{NP_k} (-1,0,1)$\\
3 & $\beta P(k)i_k \sum_{k' \in K} {P(k'|k) r_{k'}}$ & $\frac{1}{NP_k}(-1,1,0)$\\
4 & $\gamma P(k)i_k \sum_{k' \in K} {P(k'|k) \theta_{k'}}$ & $\frac{1}{NP_k} (-1,0,1)$\\ \hline
\end{tabular}
\end{center}
\caption{Probability and effect on $x_k$ for different cases}
\label{tab:tran1}
\end{table}

We now make the dependence on the population size $N$ explicit and denote by $F^N({\bf x}):=[F^N_k({\bf x})]_{k=1}^K$ the drift of ${\bf x}$. $F^N_k({\bf x})$ is as follows.

\begin{eqnarray*}
\label{eqn:drift}
%\scriptstyle{F^N_k({\bf x})}=\scriptstyle{\frac{1}{N}} 
\frac{1}{N} \left(\begin{array}{c}
 \scriptstyle{- \beta i_k \sum_{k' \in K} {P(k'|k) r_{k'}} - \gamma i_k \sum_{k' \in K} {P(k'|k) \theta_{k'}} - (\alpha + \delta) i_k}\\ 
  \scriptstyle{\beta i_k \sum_{k' \in K} {P(k'|k) r_{k'}} + \alpha i_k} \\
 \scriptstyle{\gamma i_k \sum_{k' \in K} {P(k'|k) \theta_{k'}} + \delta i_k} 
\end{array}\right)
\end{eqnarray*}

Observe that $(i)$ the number of transitions per agent per time slot is of the order of $\frac{1}{N}$ $(ii)$ the second moment of number of agent transitions per time slot is bounded and $(iii)$ $F^N({\bf x})$ is a smooth function of $\frac{1}{N}$ and ${\bf x}$. Let $F({\bf x}) = \lim_{N \to \infty} \frac{F^N({\bf x})}{1/N}$.
It then follows from Theorem 1 of \cite{Benaim08} that the time evolution of ${\bf x}(n)$ can be approximated by the following system of ODEs (with the same initial conditions). 
\begin{eqnarray}
\label{eqn:ode}
{\dot {\bf x}} = F({\bf x})
\end{eqnarray}

More explicitly, for $1 \le k \le K$
\begin{eqnarray*}
{\dot i}_k &=&  - \beta i_k R_k  - \gamma i_k \Theta_k -(\alpha+\delta) i_k \\
{\dot r}_k &=&  \beta i_k R_k + \alpha i_k \\
{\dot \theta}_k &=&  \gamma i_k \Theta_k + \delta i_k 
% {\dot i}_k &=&  - \beta i_k \sum_{k' \in K} {P(k'|k)r_{k'}}  - \gamma i_k \sum_{ k' \in K} {P(k'|k)\theta_{k'}} -(\alpha+\delta) i_k \\
% {\dot r}_k &=&  \beta i_k \sum_{k' \in K} {P(k'|k)r_{k'}} + \alpha i_k \\
% {\dot \theta}_k &=&  \gamma i_k \sum_{ k' \in K} {P(k'|k)\theta_{k'}} + \delta i_k 
\end{eqnarray*}
where, $R_k:=\sum_{k' \in K} P(k'|k)r_{k'}$ and $\Theta_k:=\sum_{ k' \in K} P(k'|k)\theta_{k'}$.

The seller offers direct incentives and referral rewards to increase $\alpha$ and $\beta$ respectively. We model this as follows. A referral reward of $c$ results in an increase of $\epsilon _1$ in $\beta$ and a direct incentive of $c'$ causes $\alpha$ to increase by $\epsilon _2$. Thus, for the duration of the referral reward program, social influence rate of $(\beta+\epsilon _1)$ is operational and the seller incurs a cost of $c$ for every successful referral. Similarly, if the direct incentive program is executed for some duration, the take-rate for seller's product increases to $(\alpha+\epsilon_2)$ for that duration, incurring her a cost of $c'$ for every sale. We normalize $c$ and $c'$ with respect to the product price. Thus the price is fixed to $1$. The seller's problem of maximizing her profit (revenue minus cost) over a fixed time horizon $T$ by optimally timing the two program can now be stated formally as follows.

Let $u_k(t)$ (resp. $v_k(t)$) denote the control variable indicating whether or not the referral reward program (resp. direct incentive program) is offered to class-k at time $t$. The cost incurred in running the referral reward program is 
\begin{eqnarray}
\label{eqn:rr}
\int_0^T \sum_{k=1}^K P(k) u_k(t)  c (\beta+\epsilon_1) i_k(t) R_k(t) dt
\end{eqnarray}
Recall that the {\em conversion rate} of potential buyers under the program is $c (\beta+\epsilon_1) i_k(t) R_k(t)$.
The cost incurred in the direct incentives program is 
\begin{eqnarray}
\label{eqn:di}
\int_0^T \sum_{k=1}^K P(k) v_k(t)  c'(\alpha+\epsilon_2) i_k(t) dt
\end{eqnarray}
Since the product price is unity, the revenue obtained is proportional to the number of customers at the end of horizon, $\sum_{k=1}^K P(k)r_k(T)$. Denoting the total cost (\ref{eqn:rr})+(\ref{eqn:di}) by $C(T)$ the problem is
\[ \mathrm{Maximize}\,\, \sum_{k=1}^K P(k)r_k(T) - C(T) \]
subject to
\begin{eqnarray*}
{\dot i}_k &=&  - (\beta+u_k \epsilon_1) i_k R_k  - (\alpha+v_k \epsilon_2) i_k - \gamma i_k \Theta_k -\delta i_k \\
{\dot r}_k &=&  (\beta+u_k \epsilon_1) i_k R_k + (\alpha+v_k \epsilon_2) i_k \\
{\dot \theta}_k &=&  \gamma i_k \Theta_k + \delta i_k 
\end{eqnarray*}
for $1 \le k \le K$ and a given initial condition ${\bf x}(0)$.

Three remarks are in order. The assumption of heterogeneity only in the number of social contacts is mainly to keep the formulation simple and highlight the impact of network structure. Extending this formulation to a general setting is straightforward and will be taken up in a longer version of the paper. Our random interaction model essentially means that the social influence on a potential buyer is the average influence from her neighbors. This, we believe, is reasonable since we have also assumed presence of external influence (through $\alpha$) on agents\footnote{For the lack of clear empirical evidence, one may also consider {\em total} influence from the neighbors. Mathematically, it is a simple modification to our formulation.}. In the above formulation we consider fixed rewards and incentives pay-outs ($c$ and $c'$). This simplifies implementation in practice. Calibration of $c$ and $c'$ can be carried out through numerical studies.

\section{Structure of Optimal Control}
\label{sec:referral}
In this section we mathematically prove the structural properties of an optimal control. To keep the proof simple, we will assume that the network $\cal{G}$ is drawn randomly from a set of regular networks of size $N$ and degree $k$. This is without loss of generality.

Let $i(t)$, $r(t)$ and $\theta(t)$ denote the fraction of population in states $\{0,1,-1\}$ at time $t$ respectively.
Let $u(t) \in \{0,1\}$ denote whether or not the referral reward program is offered at time $t$ and let $v(t) \in \{0,1\}$ denote whether or not the direct incentive program is offered at time $t$. The purchase dynamics under the influence of these programs are given as follows: 
\begin{eqnarray}
\label{eqn:i}
{\dot i}&=&-(\beta +u\epsilon _1) ir-(\alpha +v\epsilon _2) i-\gamma i \theta-\delta i \\
\label{eqn:r}
{\dot r}&=&(\beta +u\epsilon _1) ir+ (\alpha +v\epsilon _2) i \\
\label{eqn:o}
{\dot \theta}&=&\gamma i\theta+\delta i
\end{eqnarray}

From (\ref{eqn:i}), (\ref{eqn:r}), and (\ref{eqn:o}), observe that $\dot{i}+\dot{r}+\dot{\theta}=0$. Therefore, it suffices to consider any two equations. Let $\Omega:=\{(i,r)|i+r \le 1, i \ge 0, r \ge 0 \}$. Let ${\bf x}(t):=(i(t),r(t)) \in \Omega$ denote the state variable. 

The optimal control problem in this simpler setting is as follows.
\begin{eqnarray}
\nonumber
\mathrm{Maximize}\,\, && r(T)- \int_0^T cu(t)(\beta + \epsilon _1)i(t)r(t) dt \\
\label{eqn:obj}
&& - \int_0^T c'v(t)(\alpha + \epsilon _2)i(t)) dt
\end{eqnarray}
\noindent subject to (\ref{eqn:i}), (\ref{eqn:r}) and the following constraints on state and control variables: for all $0 \le t \le T$, ${\bf x}(t) \in \Omega$, $u(t) \in \{0,1\}$ and $v(t) \in \{0,1\}$.

Our main result is given in Proposition \ref{prop:mixedscheme}. It shows that an optimal strategy for the seller is to deploy the two incentive programs for at most two distinct time periods. 
\begin{proposition}
\label{prop:mixedscheme}
\begin{enumerate}
\item There exist $\tau_1, \tau_2$ ($0 \le \tau_1 \le \tau_2 \le T$) such that $u^*(t)=0$ for $\tau_1 < t \le \tau_2$ and $u^*(t)=1$ else.
\item There exist $\tau_3, \tau_4$ ($0 \le \tau_3 \le \tau_4 \le T$) such that $v^*(t)=0$ for $\tau_3 < t \le \tau_4$ and $v^*(t)=1$ else.
\end{enumerate}
\end{proposition}

\proof 
$i(0)>0$ otherwise there is no problem to solve. Observe that $\Omega$ is positively invariant. Therefore a solution starting from any initial point ${\bf x}(0)\in \Omega$ remains confined to $\Omega$. This allows us to disregard state constraints from the control formulation. 

Let $u(t),v(t) \in [0,1]$ for all $t \in [0,T]$ (This relaxation allows us to establish existence of an optimal control. We show that the optimal controls are indeed `bang-bang', i.e., $u^*(t),v^*(t) \in \{0,1\}$ for all $t$). Writing the problem in Mayer form, it can be seen that the state space (appropriately expanded with additional variables) is bounded and positively invariant (thus, state trajectories remain bounded for all admissible pairs); and the system is affine in controls (see (\ref{eqn:obj}), (\ref{eqn:i}) and (\ref{eqn:r})). Existence of an optimal control is now established by Filippov-Cesari theorem. 

From (\ref{eqn:i}), (\ref{eqn:r}), and (\ref{eqn:obj}), the Hamiltonian is written as follows.
\begin{eqnarray}
\nonumber
H({\bf x},{\bf p},u, v) &=& -c u(\beta+\epsilon _1)ir -c'v(\alpha+\epsilon _2)i \\
\nonumber
&&- p_1[(\beta+ u \epsilon _1) ir+(\alpha+v\epsilon _2)i + \gamma i\theta + \delta i ] \\
\label{eqn:refham}
&&+ p_2[(\beta +u\epsilon _1)ir + (\alpha+v\epsilon _2)i]
\end{eqnarray}
${\bf p}:=(p_1,p_2)$ denotes co-state variables. Then according to Pontryagin's Maximum Principle, there exist continuous and piecewise continuously differentiable co-state functions $p_1$ and $p_2$ that satisfy  
\begin{eqnarray}
\nonumber
{\dot p_1} &=&  -\frac{\partial H}{\partial i} \\
\nonumber
&=&[c\beta - (p_2-p_1-c)\epsilon _1]ru + [c'\alpha - (p_2-p_1-c')\epsilon _2]v \\
\label{eqn:refp1}
&&+ (p_1-p_2)(\beta r+\alpha) + p_1(\gamma(1-2i-r)+\delta)\\
\nonumber
{\dot p_2} &=&  -\frac{\partial H}{\partial r}\\
\label{eqn:refp2}
&=&[c\beta-(p_2-p_1-c)\epsilon _1)]ui+(p_1-p_2)\beta i-p_1 \gamma i,
\end{eqnarray}

% \begin{eqnarray}
% \nonumber
% {\dot p_1} &=& [c\beta - (p_2-p_1-c)\epsilon _1]ru + [c'\alpha - (p_2-p_1-c')\epsilon _2]v \\
% \label{eqn:refp1}
% &&+ (p_1-p_2)(\beta r+\alpha) + p_1(\gamma(1-2i-r)+\delta)\\
% \label{eqn:refp2}
% {\dot p_2} &=& [c\beta-(p_2-p_1-c)\epsilon _1)]ui+(p_1-p_2)\beta i-p_1 \gamma i,
% \end{eqnarray}
%
at all $t \in [0,T]$ where $u$ and $v$ are continuous and satisfy the following transversality condition %(\ref{eqn:reftrans}).
\begin{eqnarray}
\label{eqn:reftrans}
p_1^*(T)=0,\,\,p_2^*(T)=1.
\end{eqnarray}
 and also satisfy, for all $t \in [0,T]$, $u(t) \in [0,1]$ and $v(t) \in [0,1]$, 
\begin{eqnarray}
\label{eqn:hmax2}
\begin{array}{l}
H(x^*(t),p^*(t),u^*(t), v(t)) \ge H(x^*(t),p^*(t),u(t),v(t))\\
H(x^*(t),p^*(t),u(t), v^*(t)) \ge H(x^*(t),p^*(t),u(t),v(t)).
\end{array}
\end{eqnarray}

From (\ref{eqn:refham}) and (\ref{eqn:hmax2}), we get the following form for controls.
\begin{eqnarray}
\label{eqn:refu}
u^*(t) &=& \left\{\begin{array}{ll}
				1 & \mathrm{if}\,\, (p_2^*(t)-p_1^*(t)-c)\epsilon _1 > c\beta \\  
				0 & \mathrm{if}\,\, (p_2^*(t)-p_1^*(t)-c)\epsilon _1 < c\beta
					\end{array}\right.\\
\label{eqn:refv}
v^*(t) &=& \left\{\begin{array}{ll}
				1 & \mathrm{if}\,\, (p_2^*(t)-p_1^*(t)-c')\epsilon _2 > c'\alpha \\  
				0 & \mathrm{if}\,\, (p_2^*(t)-p_1^*(t)-c')\epsilon _2 < c'\alpha
					\end{array}\right.					
\end{eqnarray}
In case of equality in the conditions specified in equations (\ref{eqn:refu}) and (\ref{eqn:refv}), $u^*(t)$ and $v^*(t)$  may take any arbitrary values in $[0,1]$.

Let $\phi(t):= (p_2^*(t)-p_1^*(t)-c)\epsilon _1 - c\beta$ and
$\psi(t) := (p_2^*(t)-p_1^*(t)-c')\epsilon _2 - c'\alpha$.

We denote by $H^*_t$ the Hamiltonian along optimal state-control trajectory at time $t$. The following lemma proves that Hamiltonian will always remain positive.
\begin{lemma}
\label{lem:refhampos}
$H^*_t > 0$ $\forall$ $t \in [0,T]$.
\end{lemma}

\proof From (\ref{eqn:refham}) and (\ref{eqn:reftrans}), we have
\begin{eqnarray*}
H^*_T &=& [(1-c)\epsilon _1-c\beta ]i^*(T) r^*(T) u^*(T) \\
&&+ [(1-c')\epsilon _2-c'\alpha ]i^*(T) v^*(T) \\
&&+ (\beta i^*(T) r^*(T) + \alpha i^*(T))
% H^*_T &=& [(1-c)\epsilon _1-c\beta ]i^*(T) r^*(T) u^*(T) + [(1-c')\epsilon _2-c'\alpha ]i^*(T) v^*(T) \\
% &&+ (\beta i^*(T) r^*(T) + \alpha i^*(T))
\end{eqnarray*}
$r(t)$ is non-decreasing whereas $i(t) \downarrow 0$ and $i(t) > 0$ for all $t$ since $i(0) > 0$. Therefore, $H^*_T > 0$. The conclusion follows by noting that the Hamiltonian is constant for autonomous systems. \qed %\hfill{$\Box$}%\hfill{\Halmos}
\endproof

The lemma below shows that the co-state variables remain positive for the whole duration.
\begin{lemma}
\label{lem:refppos}
$p_1^*(t), p_2^*(t) > 0$ $\forall$ $t \in [0,T)$.
\end{lemma}
\proof  Suppose $p_1^*(t) \le 0$ for all $t$ and let $p_1^*(t)=0$ at $t=\tau$ (at least one $\tau$ exists since $p_1^*(T)=0$). Then $p_2^*(\tau)>0$ otherwise $H_{\tau}^* < 0$ since $u^*(\tau)=0$. Observe from (\ref{eqn:refp1}) that ${\dot p}_1 < 0$ if $p_1=0$. Strict inequality in (\ref{eqn:refu}) implies that at $\tau$, $u^*(\cdot)$ is continuous. Therefore, ${\dot p}_1 < 0$ in the neighborhood of $0$. Thus $p_1^*(t_1) \le 0$ implies $p_1^*(t) < 0$ for all $t>t_1$ and $p_1^*(T) \neq 0$ which violates (\ref{eqn:reftrans}). It follows that $p_1^*(t) > 0$ for all $t \in [0,T)$. This in turn implies that $p_2^*(t) > 0$ for all $t$ otherwise $H_t^* < 0$. \qed % \hfill{$\Box$} 
% since $u^*(t)=0$. 
\endproof

\begin{lemma}
\label{lem:refpgreat}
$p_2^*(t)>p_1^*(t)$ $\forall$ $t \in [0,T]$.
\end{lemma}
\proof Suppose not. Let $p_2^*(t)<p_1^*(t)$ at $t=\tau$. Then $\phi(\tau)<0$ and, therefore, $u^*(\tau)=0$. (\ref{eqn:refham}) then yields $H^*_{\tau} < 0$, a contradiction. \qed %\hfill{$\Box$}
\endproof
Let  $\zeta(t):= (p_2^*(t)-p_1^*(t))i^*(t)$. The lemma that follows shows that $\zeta(t)$ is a decreasing function.
\begin{lemma}
\label{lem:psidot}
${\dot \zeta}(t) < 0$ $\forall$ $t \in [0,T]$.
\end{lemma}
\proof From (\ref{eqn:i}), (\ref{eqn:refp1}), and (\ref{eqn:refp2}) we get
\begin{eqnarray*}
{\dot \zeta}(t) &=& -[c(\epsilon _1 +\beta) u^*(t) r^*(t) + \phi(t) u^*(t) i^*(t)  \\
&& + c'(\epsilon _2 +\alpha )v^*(t)\\
&& + p_2^*(t)(\gamma (1-i^*(t)-r^*(t))+\delta)] i^*(t)
%{\dot \psi}(t) &=& -[\phi(t) \zeta u^*(t) i^*(t) + c (\beta+\zeta) u^*(t) r^*(t) + (p_2^*(t)-p_1^*(t)) \beta+p_2^*(t) (\gamma(1-i^*(t)-r^*(t))+\delta)] i^*(t)
\end{eqnarray*}
Lemma follows by noting that all terms inside the bracket are non-negative. \qed %\hfill{$\Box$}
\endproof
Now consider ${\dot \phi}(t)$. From (\ref{eqn:refp1}), (\ref{eqn:refp2}), and (\ref{eqn:refham}) we get
\begin{eqnarray*}
{\dot \phi}(t) &=& [\frac{H^*_t}{i^*(t)} - \phi(t) i^*(t) u^*(t) - (p_2^*(t)-p_1^*(t))\beta i^*(t)]\epsilon _1
\end{eqnarray*}
$(p_2^*(t)-p_1^*(t)) i^*(t)$ is monotonically decreasing (Lemma~\ref{lem:psidot}). From (\ref{eqn:i})
$i^*(t) \downarrow 0$ exponentially ($i^*(t)<i^*(0)e^{-(\alpha+\delta)t}$). $H^*_t$ is a positive constant (Lemma~\ref{lem:refhampos}). 

Assume that $\phi(t)=0$ at three points in time $\tau _1$, $\tau _2$, $\tau _3$. Therefore, ${\dot \phi}(\tau) >0$  for either $\tau = \tau _1$ or $\tau = \tau_2$. Without loss of generality, let us say ${\dot \phi}(\tau _2) > 0$. From the above equation it follows that ${\dot \phi}(\tau _3) > 0$ which is not feasible as $\phi(\tau_3^-)>0$. It follows that $\phi(t)=0$ at at most  two points in time. Therefore, there exist $0 \le \tau_1 \le \tau_2 \le T$ such that $u^*(t)=1$ for $0 \le t \le \tau_1$ and $\tau_2 < t \le T$, and $0$ elsewhere. 

Similarly, one can show that there exist $0 \le \tau_3 \le \tau_4 \le T$ such that $v^*(t)=0$ for $\tau_3 < t \le \tau_4$ and $v^*(t)=1$ otherwise. The proposition is, thus, established. \hfill \QED

\endproof

%
%Thus the optimal controls $u^*(t)$ and $v^*(t)$ exihibit switching behavior with at most one upward switch. This gives a simple and elegant optimal strategy structure which makes it easy to implement for the seller. %Also, given the simple structure of the optimal strategy, one can carry out numerical optimization over such strategies to compute $\tau$'s. 

Proposition~\ref{prop:mixedscheme} implies that both the referral reward and direct incentives programs are to be deployed at most twice for certain durations, one in the beginning and the other at the end. It may happen that both the durations are of length $0$ which means that a program is not deployed at all. On the other hand, it could also get deployed over the complete time horizon $T$. This gives a simple and elegant marketing strategy which is easy to implement for the seller. 
 
The structure of the above optimal control is quite intuitive. In the case of the referral reward program, the cost is proportional to the product of number of potential buyers and customers. Hence to keep the cost low, rewards are declared in the initial stage (when the number of customers is less) to motivate product advocates and  may also be paid at the end (when the number of potential buyers is less) to acquire some additional customers.

In the case of direct incentives, the cost is proportional to the number of potential buyers. If the initial take rate for the product is less, this program may get executed at initial stages to quickly acquire customers whose social influence can be exploited in the later stages; otherwise more agents may buy competitor's product and attract other potential buyers. Towards the end of the campaign, the number of potential buyers is less; hence direct incentives may be offered to attract additional customers. 

\section{Numerical Results}
\label{sec:numerical}

The simple structure of optimal controls given by Proposition~\ref{prop:mixedscheme} allows one to devise incentives programs
quite easily by numerical optimization of $\tau$'s. Here we obtain an independent validation of optimal controls by discretizing (\ref{eqn:obj}), casting it as a nonlinear constrained optimization problem and using a gradient descent approach to find an optimal solution. (For discussion on various numerical solution techniques for such problems refer to \cite{Kirk70}).  
For all our experiments, the time horizon $T$ and discretization step-size are fixed at 10 and 0.1 respectively. 
The NLP formulation is not convex. Therefore, we use a multi-start mechanism to determine an optimal solution. 
Results are also verified using the commercial package PROPT which uses pseudospectral methods for solving such problems. 

In this paper, we will primarily investigate the initial condition $i(0)=1$. This captures the case when the seller and the competitor(s) enter the market with substitutable products at around the same time (e.g., gaming technologies) or when the seller introduces an independent product into the market (e.g., a book). Of course, similar results can be obtained for the case where seller and/or competitor already have some presence in the market ($r(0)>0$ and/or $\theta (0)>0$).

\begin{figure}
\centering
\includegraphics[width=3.4in, height=2in]{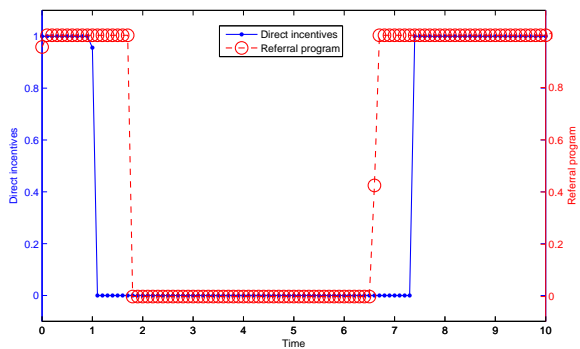}
\caption{Optimal marketing strategy for the base scenario}
\label{fig:inc6}
\end{figure}

We fix $\epsilon _1 = \epsilon _2 = 0.05, \gamma = 0.1$, $\delta =0.1$ for all numerical studies, and consider $\beta =0.1, \alpha =0.08, c=0.25$ and $c'=0.3$ as the {\em base scenario} (These parameter values are arbitrary and only roughly based on some available data).  The optimal marketing strategy is shown in Figure~\ref{fig:inc6}. It is optimal for the seller to run both the incentive programs initially for some duration, stop and then run the programs again towards the end. Note that the optimal strategy is open loop; hence estimates of $i$ and $r$ are not required for implementation.

It is, thus, possible for the seller to determine the timing of her incentives programs numerically. Experimentation with different values of pay-outs $c$ and $c'$ (which essentially fix $\epsilon_1$ and $\epsilon_2$) can be used to understand trade-offs and optimize these pay-outs. 

In the following we undertake an investigation of two important questions pertaining to the interplay between the two incentives programs and the impact of network structure on the them. The former question is important because  {\em influence-and-exploit} strategy has received much attention in the literature. As we show below, {\em exploit-and-influence} strategy can also come into play for some parameter settings. The second question is linked to the so-called {\em influentials hypothesis} which informally says that high degree agents ({\em hubs}) play significant role in product diffusion, and, therefore, are natural targets for incentives (direct or referral rewards). We show that in some cases the seller is better off incentivizing low degree agents (more than high-degree ones). Thus, our results highlight the need for a careful consideration of the network structure while making incentive decisions.

%\cite{Watts07} reports that in most cases large cascade of influence is not driven by high degee nodes (they refer to them as influentials or hubs). On the other hand, \cite{Goldenberg09} highlights the importance of high degree nodes (also called influentials or hubs) in adoption process. In Section \ref{sec:impact}, we examine the impact of network structure on incentive strategies and show that for some structures the seller may be better off incentivizing mostly low degree nodes whereas for some other structure she targets high degree nodes.  

\subsection{Interplay between Referral and Direct Incentive Programs}
\label{sec:interplay}
\begin{figure}[t]%[hbt!]
\begin{minipage}{3.4in}
\centering
\includegraphics[width=3.4in, height=2in]{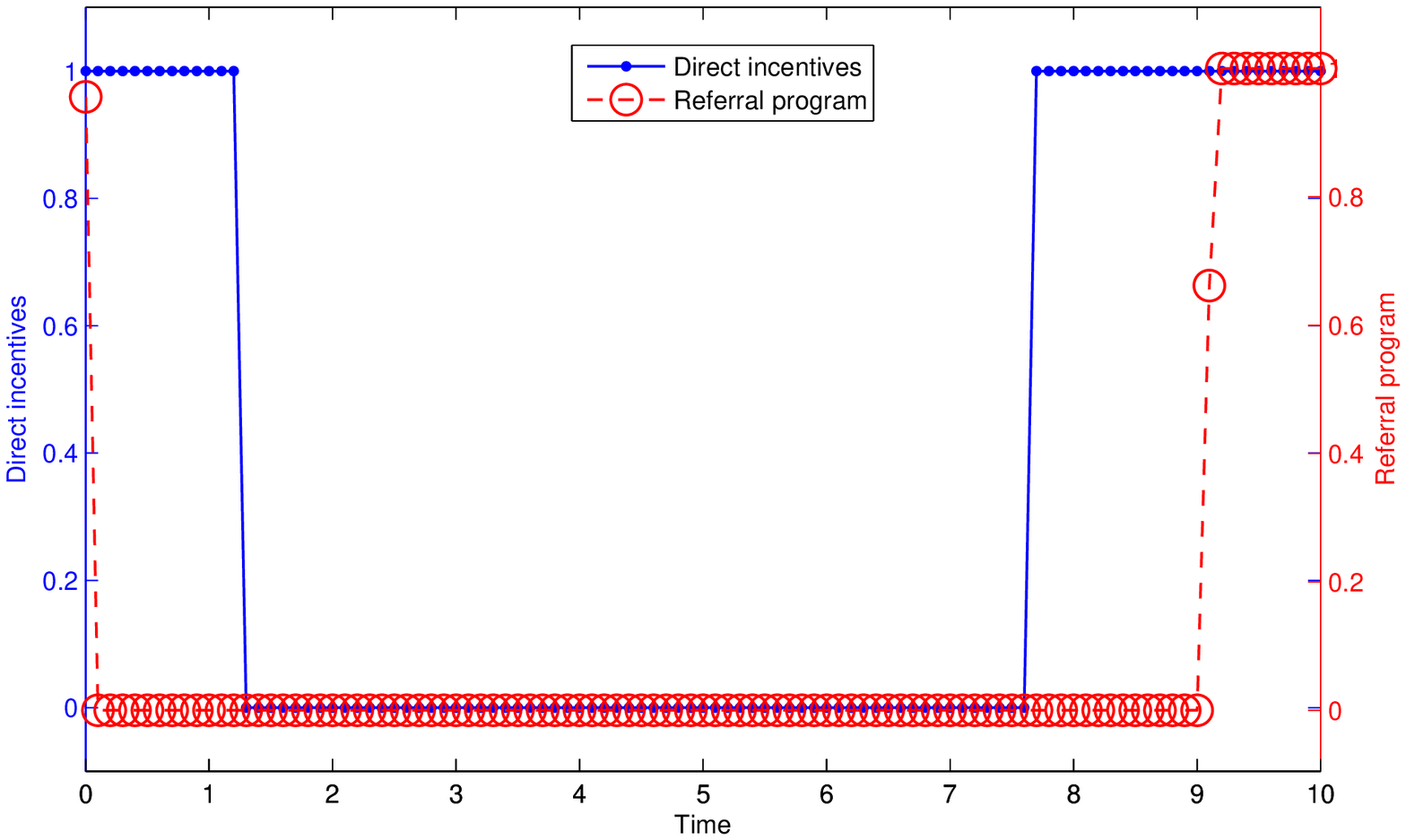}
\caption{{\em Influence-and-exploit} strategy is optimal when $\beta$ is increased to $0.13$}
\label{fig:inc11}
\end{minipage}
\hfill
\begin{minipage}{3.4in}
\centering
\includegraphics[width=3.4in, height=2in]{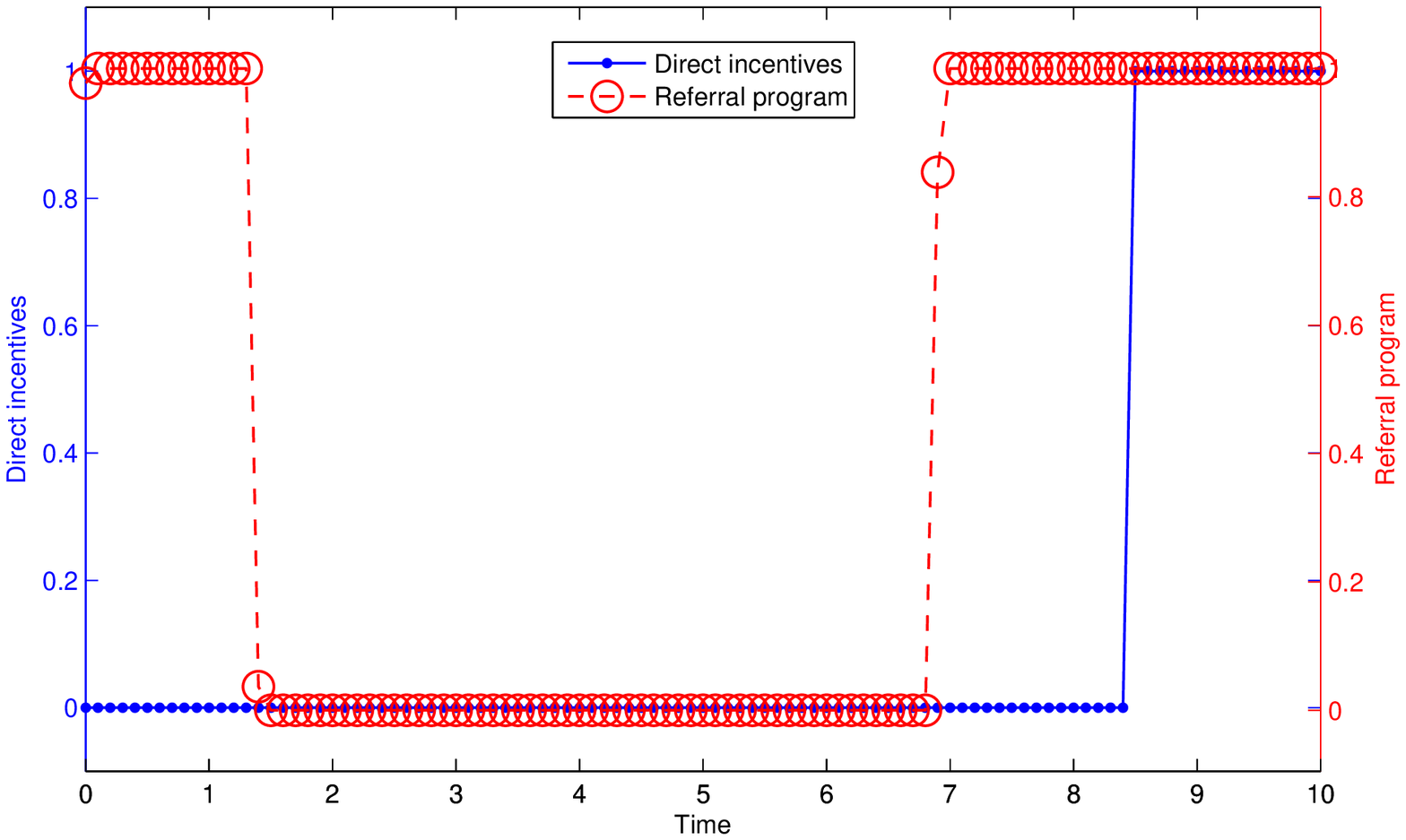}
\caption{{\em Exploit-and-influence} strategy is optimal when $\alpha$ is increased to $0.09$}
\label{fig:inc10}
\end{minipage}
\end{figure}

When social influence is strong in the population, i.e., $\beta$ is higher, the seller needs to employ only direct incentives initially. For example, if $\beta$ is set to $0.13$ in the base scenario then it is optimal to offer referral rewards only at the end and that too for a short period as shown in Figure~\ref{fig:inc11}. This can be seen as a manifestation of the {\em influence-and-exploit} strategy. On the other hand, if the seller has established a good reputation in the market, translating into a higher value of $\alpha$, an initial influence step through direct incentives may not even be required. See from Figure~\ref{fig:inc10} that when $\alpha =0.09$ in the base scenario, it is optimal for the  seller to offer direct incentives only at the end. In this case, for most portion of the time horizon, the seller must exploit connections of existing customers and only at the end must she impart direct influence on potential buyers. We call it the {\em exploit-and-influence} strategy for the seller.

\begin{figure}[t]
\begin{minipage}{3.4in}
\centering
\includegraphics[width=3.4in, height=2in]{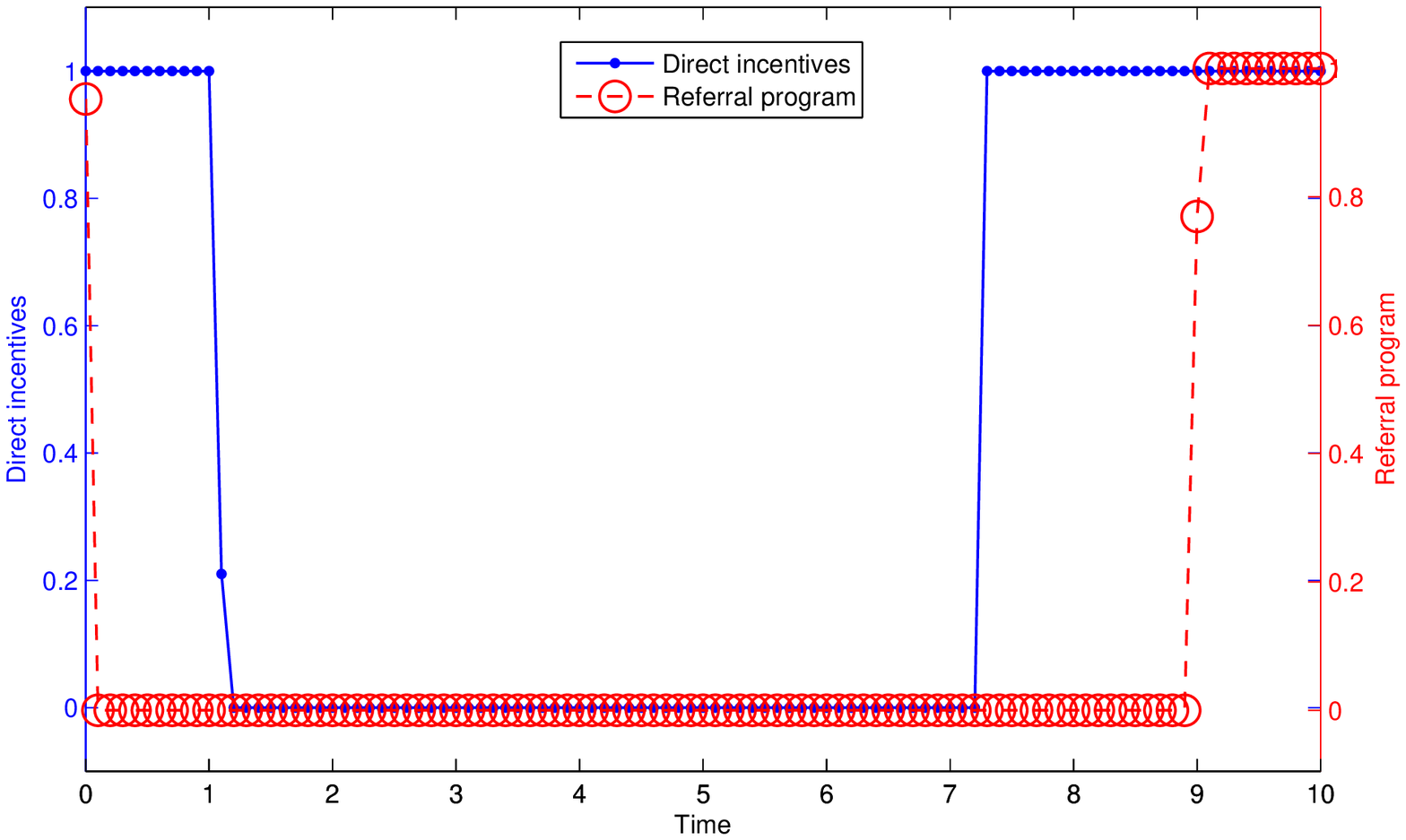}
\caption{{\em Influence-and-exploit} strategy is optimal with pay-outs: $c=0.3$, $c'=0.3$} 
\label{fig:inc7}
\end{minipage}
\hfill
\begin{minipage}{3.4in}
\centering
\includegraphics[width=3.4in, height=2in]{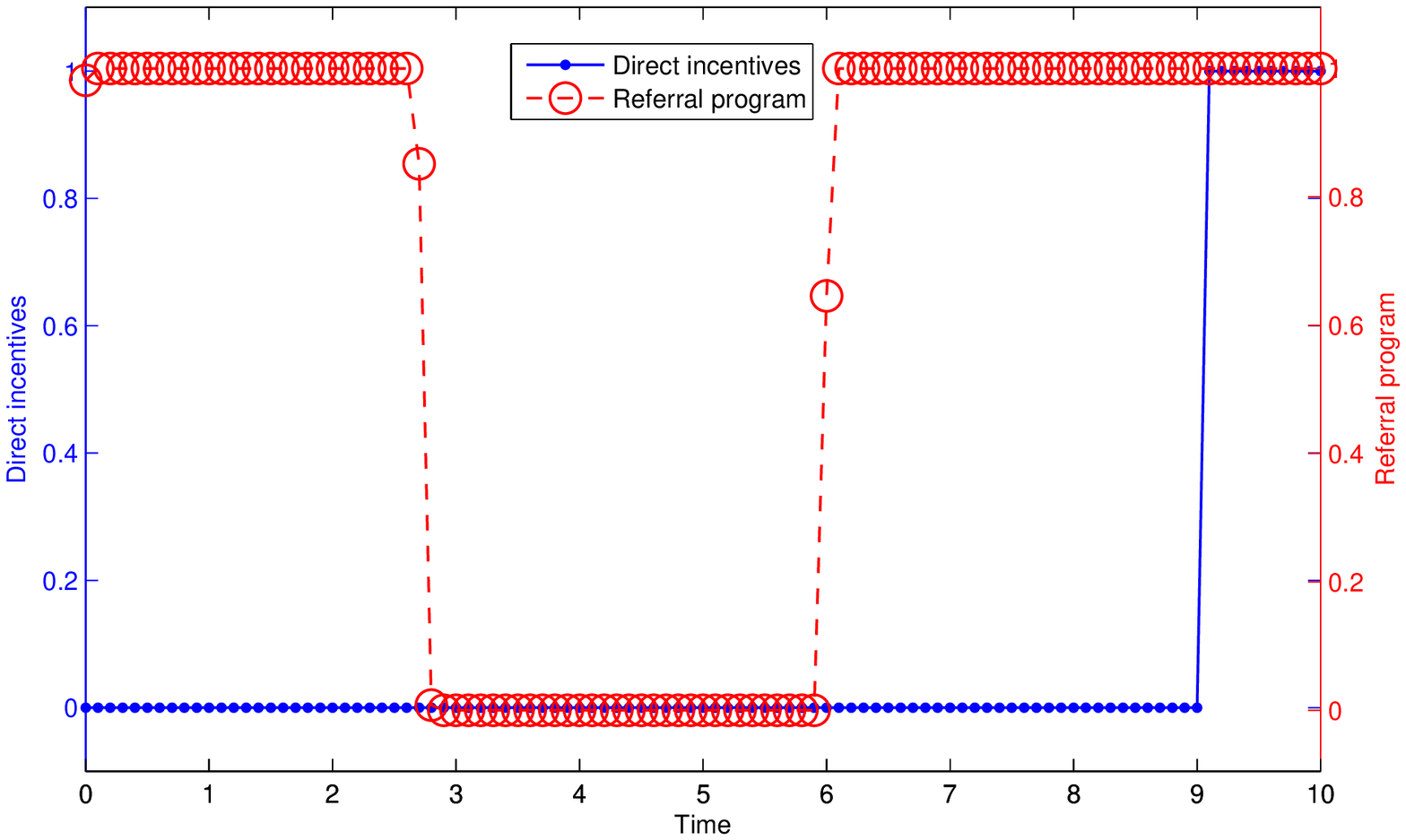}
\caption{{\em Exploit-and-influence} strategy is optimal with pay-outs: $c=0.25$, $c'=0.35$}
\label{fig:inc5}
\end{minipage}
\end{figure}

We also observe from Figures~\ref{fig:inc7} and \ref{fig:inc5} that {\em influence-and-exploit} and {\em exploit-and-influence} are optimal strategies for the seller if she incurs high per conversion pay-outs for referral and incentive programs respectively.

\subsection{Impact of Network Structure on Incentive Programs}
\label{sec:impact}
%Real world networks show strong degree correlation amongst connected nodes. Some networks show assortative mixing of nodes by degrees where high-degree nodes have most of their connections to other high-degree nodes. Others show disassortative mixing where high-degree nodes have most of their connections to low-degree nodes \cite{Newman02}.

Real-world networks show strong degree correlation amongst connected nodes. Some networks show {\em assortative} mixing of nodes by degrees where high-degree nodes have most of their connections to other high-degree nodes. Others show {\em disassortative} mixing where high-degree nodes have most of their connections to low-degree nodes (\cite{Newman02}). In this section, we examine the impact of network structure on incentives programs.

We consider an undirected correlated network with nodes belonging to either of two classes $A$ and $B$ with probability $P(A)$ and $P(B)$ respectively. Class $A$ nodes are of high degree, say $k_A$ and class $B$ nodes are of low degree, say $k_B$. $P(A|B)$ is the probability that a given link from class $B$ node points to a class $A$ node. $P(B|A)$ can be computed from the following balance equation:
\begin{eqnarray}
\label{eqn:balance}
k_A P(B|A) P(A) = k_B P(A|B) P(B)
\end{eqnarray}  

We consider two types of network structures. One structure represents assortative mixing whereas the other one represents disassortative mixing. To keep things simple and derive key insights, we assume that the seller is optimizing implementation of only one incentives program, namely, referral rewards program. The seller can offer referral rewards to class $A$ and/or class $B$ nodes to increase their social influence rate ($\beta$)  by $\epsilon$. As earlier, she incurs a per conversion cost of $c$ after normalizing with respect to product price.

We fix $\alpha = 0.1, \delta =0.1, \beta = 0.1, \gamma=0.15, \epsilon=0.08, k_A=10, k_B=2, P(A)=0.1, P(B)=0.9$ for our numerical studies. For disassortative network, we set $P(B|A) = 0.9$ whereas for assortative network, we set $P(B|A)=0.1$. 

\begin{figure}[t]
\begin{minipage}{3.4in}
\centering
\includegraphics[width=3.4in, height=2in]{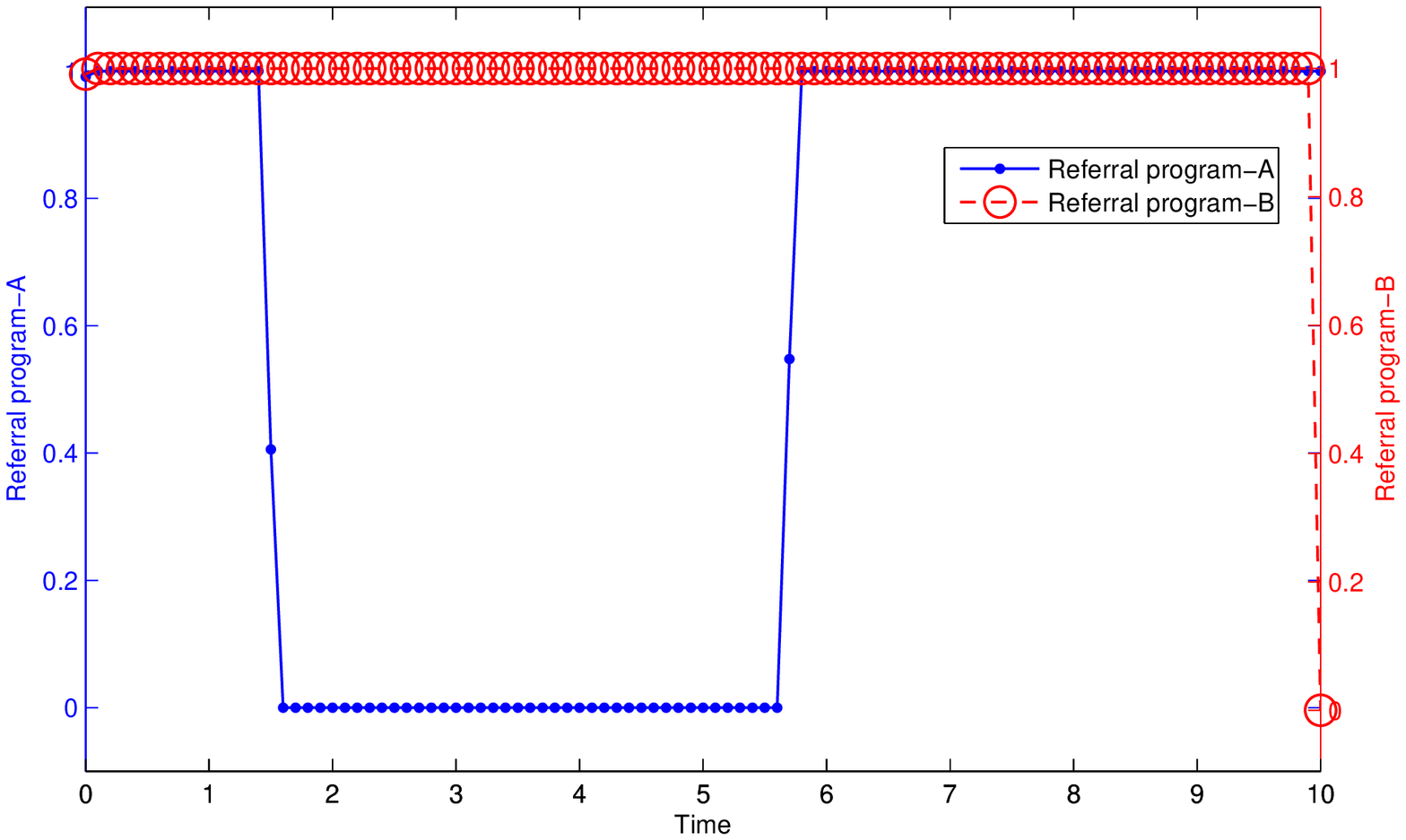}
\caption{Optimal marketing strategy on a network with disassortative mixing}
\label{fig:incdis}
\end{minipage}
\hfill
\begin{minipage}{3.4in}
\centering
\includegraphics[width=3.4in, height=2in]{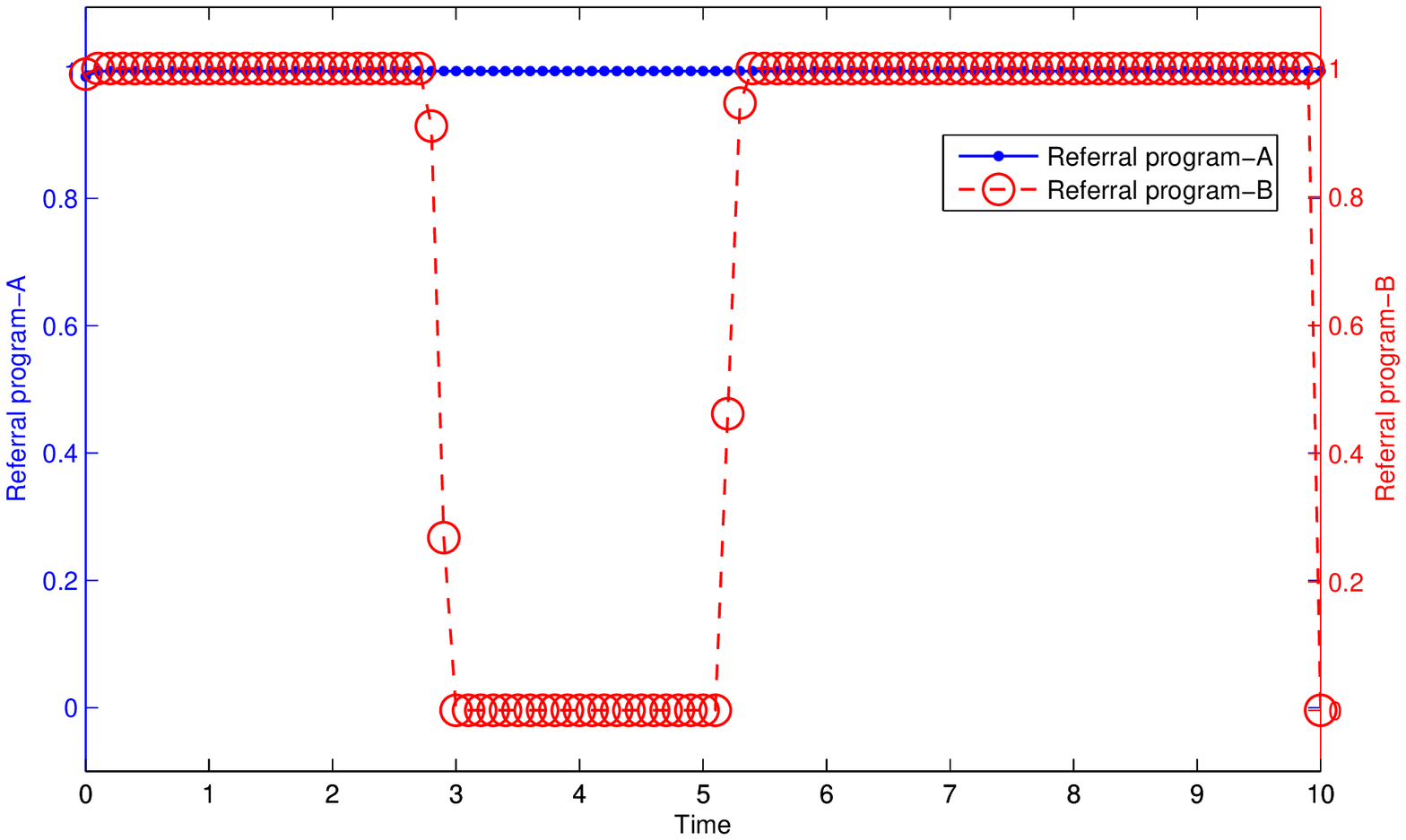}
\caption{Optimal marketing strategy on a network with assortative mixing}
\label{fig:incass}
\end{minipage}
\end{figure}

The optimal timing of referral reward program for disassortative network is shown in Figure~\ref{fig:incdis}. The seller's optimal strategy is to offer referral rewards to class $B$ nodes for the complete duration whereas rewards to class $A$  nodes are offered initially for a short duration and then again towards the end for a short duration. 
In this case, class B nodes have almost half of their connections going to class A nodes. Also, major fraction of the population is from class B.  So, referral rewards are offered to class B nodes for entire duration as it increases influence not only on class B nodes but also on class A nodes. Class A nodes are not rewarded for the entire duration in order to control the cost.

In the case of assortative network, the optimal strategy changes completely (see Figure~\ref{fig:incass}). The seller offers referral rewards to class $A$ nodes for the complete duration whereas rewards to class $B$ nodes are offered initially for some duration and then again towards the end. 
In this case, nodes from both the classes are well connected amongst themselves with very few connections going across the classes.  So, the optimal reward strategies for both the classes are essentially independent. For this particular scenario, it turns out that it is optimal to offer rewards to class A nodes for the entire duration as the cost incurred is not much.  Whereas in the case of class B nodes, referral rewards are discontinued for some duration in the middle as the cost overshoots the potential revenue.  

The results show that networks with different structures can result in different optimal strategies for the seller. In some scenarios, the seller may be better off incentivizing low degree nodes as against the popular approach of targeting the influentials (high degree nodes), thus, providing some support to the finding in \cite{Watts07}. In some scenarios, the seller may be better off targeting influentials thus supporting the results in \cite{Goldenberg09}. Thus, our results highlight the need for a careful consideration of the network structure while making incentive decisions.  

%We also experiment with a scenario where social influence of a node depends on her class. We assume nodes with high degree (class $A$ in our case) impart more social influence. We fix $\beta_A = 0.15, \beta_B = 0.1, \gamma_A = 0.18, \gamma_B = 0.15$. Other parameters remain unchanged. The optimal strategy for dissortative network is as shown in Figure~\ref{fig:incdisbeta15}. The result again supports ``influentials hypothesis''.  

%\cite{Watts07} reports that in most cases large cascade of influence is not driven by high degee nodes (they refer to them as influentials or hubs). On the other hand, \cite{Goldenberg09} highlights the importance of high degree nodes (also called influentials or hubs) in adoption process. In Section \ref{sec:impact}, we examine the impact of network structure on incentive strategies and show that for some structures the seller may be better off incentivizing mostly low degree nodes whereas for some other structure she targets high degree nodes.

% \begin{figure}
% \centering
% \includegraphics[width=3.4in, height=2in]{./figurescopy/incentassbeta15.eps}
% \caption{Optimal marketing strategy for class-based social influence on a network with disassortative mixing.}
% \label{fig:incdisbeta15}
% \end{figure} 

\section{Conclusion}
\label{sec:conclusion}
In this paper we have addressed the problem of optimal timing of two incentive programs, namely, direct incentives and referral rewards, for product diffusion through social networks. Taking a deviation from the existing approaches, we formulate the problem as a continuous-time deterministic optimal control problem. The optimal strategy for the seller is to deploy these programs in at most two distinct time periods. The simplicity of this structure and non-adaptive nature makes them ideal for implementation in practice. We further show that if the seller has good reputation in the market, {\em exploit-and-influence} strategy can be optimal whereas if social influence is strong in the population, {\em influence-and-exploit} strategy can be optimal for the seller.  In the case of correlated networks, our numerical studies show that the seller need not necessarily offer more frequent referral reward programs to high degree nodes to maximize her profit.

There are two immediate directions for future work: $(i)$ extend heterogeneity of agents to include their external and social influence probabilities and $(ii)$ devise procedures to estimate model parameters.

%\subsubsection*{Acknowledgements}
%The authors would like to thank K. Ravikumar for his support and valuable feedback on the paper. 

%\bibliographystyle{elsarticle-num}
%\bibliographystyle{splncs}
%\bibliographystyle{ormsv080} % outcomment this and next line in Case 1

\bibliographystyle{plainnat}
\bibliography{wine} % if more than one, comma separated

\end{document}